# Zeroth law of thermodynamics for thermalized open quantum systems having constants of motion

V. Yu. Shishkov,[1,2] E. S. Andrianov,[1,2] A. A. Pukhov,[1,2,3] A. P. Vinogradov,[1,2,3] and A. A. Lisyansky[4,5]

[1]Dukhov Research Institute for Automatics, 22 Sushchevskaya, Moscow 127055, Russia

[2]Moscow Institute of Physics and Technology, 9 Institutskiy per., Dolgoprudniy 141700, Moscow Reg., Russia

[3]Institute for Theoretical and Applied Electromagnetics RAS, 13 Izhorskaya, Moscow 125412, Russia

[4]Department of Physics, Queens College of the City University of New York, Queens, NY 11367, USA

[5]The Graduate Center of the City University of New York, New York, New York 10016, USA

We study the evolution of an open quantum system described by a dynamical semigroup having the Lindblad superoperator as a generator. This generator may have an eigenfunction with a unity eigenvalue, referred to as a constant of motion (COM). An open quantum system has a unique stationary state if and only if it has no COMs. A system with multiple stationary states has a basis of COMs; any COM of the system is a linear combination of the basis COMs. The basis divides the space of system states into subspaces. Each subspace has its own stationary state, and any stationary state of the system is a linear combination of these states. Usually, neither the basis of COMs nor even the number of COMs is known. We demonstrate that finding the stationary state of the system does not require looking for the COMs. Instead, one can construct a set of "invariant" subspaces. If the system evolution begins from one of these subspaces, the system will remain in it, arriving at a stationary state independent of evolution in other subspaces. We suggest a direct way of finding the invariant subspaces by studying the evolution of the system. We show that the sets of invariant subspaces and subspaces generated by the basis of COMs are equivalent. A stationary state of the system is a weighted sum of stationary states in each invariant subspace; the weighted factors are determined by the initial state of the system.

## I. INTRODUCTION

Recently, the applicability of the laws of thermodynamics to open quantum systems interacting with reservoirs has been actively discussed [1-13]. This issue is interesting not only from a fundamental point of view but is also important for practical purposes. Many applications require creating a system state with desired properties, e.g., quantum entanglement of a large array of qubits for quantum computer elements [14-16], antibunched photons for quantum cryptography [17,18], and a coherent state of an electromagnetic field for nanoscale radiation sources [19-22]. Attaining as well as retaining desired states of an open system is a difficult problem because the system interacts with an external reservoir, and the outcome of this interaction is constrained by the laws of thermodynamics. First, the laws of thermodynamics determine possible system states. Second, according to thermodynamics, any state should relax to the stationary state determined by coupling with the reservoir. This significantly limits the number of desirable states.

However, the applicability of the laws of thermodynamics to quantum systems is still not clear. Under the assumption that the density operators of the system and the reservoir are always separable, that the reservoir state does not change in time (the Born approximation), and that the system dynamics is local in time (the Markov approximation), one can obtain the master equation for the density matrix of the system in the Lindblad-Gorini-Kossakowski-Sudarshan (LGKS) form [23-27].

It has been shown that the first law and the second law in the Clausius form follow from this equation [6,26,28]. The zeroth law, which affirms that the stationary state of the system has a unique Gibbs distribution with the reservoir temperature (system thermalization) holds if and only if the system does not have constants of motion (COMs) [28-30].

A COM $\hat{I}(t)$ is s an eigenoperator of the evolution generator $\exp(\hat{L}t)$, which eigenvalue is equal to unity. It has been shown [30] that $\hat{I}(t)$ should be invariant under the action of the Lindblad superoperator $\hat{L}$:

$$\hat{L}(\hat{I}) = -i\left[\hat{H}_s, \hat{I}\right] + \frac{1}{2}\sum_i \left(\left[\hat{S}_i, \hat{I}\hat{S}_i^\dagger\right] + \left[\hat{S}_i\hat{I}, \hat{S}_i^\dagger\right]\right) = 0, \tag{1}$$

where $\hat{A}$ is an arbitrary operator, $[,]$ denotes a commutator, the summation is taken over eigenstates of the system, and $H_S$ is the system Hamiltonian. In the derivation of Eq. (1) it is assumed that the Hamiltonian of the interaction between the system and the reservoir is taken in the form $\hat{H}_{SR} = \hbar\lambda\hat{S}\hat{R}$ [26,31,32], where $\hat{S}$ and $\hat{R}$ are dimensionless operators that only depend on dynamical variables of the system and the reservoir, respectively. The interaction parameter $\lambda$ has the dimension of frequency. Thus, $\hat{I}(t)$ is a COM, if $\hat{L}\left[\hat{I}(t)\right] = 0$. In this case, $\hat{I}(t)$ commutes with both the system Hamiltonian and the operator $\hat{S}$ [6].

Since a COM $\hat{I}(t)$ commutes with $H_S$, these two operators have a common set of eigenvectors, referred to below as basis vectors. Following the general theory [30], we must find a basis of COMs, the linear combinations of which generates all possible COMs of the system. In Ref. [30], it has also been shown that the basis of COMs is mapped into the family of projection operators that divides the space of system state into subspaces. The existence theorem (see Ref. [30]) establishes that in each subspace, its own stationary state is formed, and that any stationary state of the system is a linear combination of these states. In other words, the determination of stationary states requires knowledge of the basis COMs. However, there are no general recipes for finding COMs or even for determining their total number [33-38].

It seems that the only way for the implementation of this highly abstract theory is to run over all possible operators to find COMs. Since a COM is diagonal in the basis of eigenvectors of $H_S$, a general form of a COM is a diagonal matrix containing $n$ ones and $N-n$ zeros that occupy arbitrary places, where $N$ is the rank of system state space. The total number of such matrices is $2^N$. To choose COMs of $2^N$ matrices, one needs to make sure that they satisfy the equation $L\left[\hat{I}(t)\right] = 0$.

The next step is to determine the basis COMs. If there is only one COM, then the states with a certain eigenvalue of this COM can be separated as a subspace. Thus, each COM leads to a division of the state space into subspaces. The division that corresponds to the basis COMs is the intersection of all subspaces of all COMs. Finally, the eigenvalues of $H_S$, which correspond to the eigenvectors belonging to one of such subspaces, determine the partition function and the Gibbs distribution in the subspace.

In this paper, we propose a way of determining stationary states of an open quantum system. The developed approach only requires knowledge of the Hamiltonians of the system and the system-reservoir interaction; it does not require knowing either COMs or their number. Moreover, the proposed method enables one to find all basis COMs. The method is based on the determination on invariant subspaces. These are such subspaces that if the system evolution begins from one of them, the system remains in this subspace reaching the stationary state. We also show that the sets of invariant and basis subspaces are equivalent. The behavior of the system inside a subspace is equivalent to the behavior of the system without COMs, and according to Ref. [30], its stationary state would be described by the Gibbs distribution. The stationary state of the whole system depends on the projection of the initial state onto the subspaces. It is a weighted sum of the stationary states in each invariant subspace. The weighted factors are determined by the initial state of the system.

## II. MASTER EQUATION FOR OPEN QUANTUM SYSTEM

Let us consider the system $S$ described by the Hamiltonian $\hat{H}_S$. Via the interaction Hamiltonian, $\hat{H}_{SR}$, the system interacts with the reservoir $R$ having the Hamiltonian $\hat{H}_R$. The dynamics of the system and the reservoir is described by the von Neumann equation for the density matrix $\rho$

$$\frac{d\rho(t)}{dt} = \frac{i}{\hbar}\left[\rho(t), \hat{H}_S + \hat{H}_R + \hat{H}_{SR}\right]. \tag{2}$$

In the Born-Markov approximation, one can eliminate the reservoir degrees of freedom and reduce Eq. (2) to the master LGKS equation which describes the dynamics of the system density matrix, $\rho_S = Tr_R \rho$, only [6,23,26,28]:

$$\frac{\partial}{\partial t}\hat{\rho}_S(t) = L[\hat{\rho}_S(t)], \tag{3}$$

where the operator $L[\hat{\rho}_S(t)]$ is determined by Eq. (1)

$$L[\hat{\rho}_S(t)] = -\frac{i}{\hbar}[\hat{H}_S, \hat{\rho}_S] + \lambda^2 \sum_{k_1,k_2} G(\omega_{k_1} - \omega_{k_2})\left(\left[\hat{S}_{k_1 k_2}, \hat{\rho}_S(t)\hat{S}^\dagger_{k_1 k_2}\right] + \left[\hat{S}_{k_1 k_2}\hat{\rho}_S(t), \hat{S}^\dagger_{k_1 k_2}\right]\right) \tag{4}$$

where the operator $\hat{S}$ is presented as a sum of the operators $\hat{S}_{k_1 k_2} = S_{k_1 k_2}|k_1\rangle\langle k_2|$,

$$\hat{S} = \sum_{k_1, k_2} S_{k_1 k_2}|k_1\rangle\langle k_2|, \tag{5}$$

where $|k_i\rangle$ ($i=1,...,N$) are non-degenerate eigenstates of the system Hamiltonian $\hat{H}_S$, $\omega_{k_i}$ are eigenfrequencies corresponding to these states, and the function

$$G(\omega) = \int_{-\infty}^{\infty} \exp(i\omega\tau)\, \text{Tr}_R\left(\hat{\tilde{R}}(t)\hat{\tilde{R}}(t+\tau)\hat{\rho}_R\right) d\tau$$

is the Fourier transform of the reservoir correlation function

$$\hat{\tilde{R}}(t) = \exp(i\hat{H}_R t/\hbar)\hat{R}\exp(-i\hat{H}_R t/\hbar).$$

Note that if the reservoir has temperature $T$, i.e.

$$\hat{\rho}_R = \exp(-\hat{H}_R/kT)/\text{Tr}\exp(-\hat{H}_R/kT),$$

then the Kubo-Martin-Schwinger condition,

$$G(\omega) = \exp(\hbar\omega/kT)G(-\omega)$$

is satisfied.

Now we show that a COM commutes with $\hat{H}_{SR}$. By definition [6], the COM, say $\hat{I}_S$, should commute with $\hat{H}_S$ and, consequently, it has a diagonal form in the eigenbasis of $\hat{H}_S$, $\hat{I}_S = \sum_k I_k |k\rangle\langle k|$. The dynamics of the expected value of the operator $\hat{I}_S$ can be obtained via the equation

$$\frac{d}{dt}\langle\hat{I}_S\rangle = Tr_S\left(\dot{\rho}_S \hat{I}_S\right) = \sum_{k_1,k_2} \gamma_{k_1,k_2} Tr\left(|S_{k_1 k_2}|^2 (I_{k_1} - I_{k_2})|k_2\rangle\langle k_2|\rho\right) = 0. \tag{6}$$

where time evolution of the density matrix is governed by Eq. (3). Equation (6) should be valid for any density matrix $\hat{\rho}$ at any moment, including the initial moment. Since at the initial moment, we can choose $\hat{\rho}$ to be an arbitrary density matrix, then Eq. (6) reduces to

$$|S_{k_1 k_2}|^2 (I_{k_1} - I_{k_2}) = 0, \tag{7}$$

for arbitrary $k_1$ and $k_2$. At the same time, the commutativity of $\hat{I}$ and $\hat{H}_S$ gives

$$\left[\hat{I}_S^*, \hat{S}\right] = \sum_k I_S^{*(k)} |k\rangle\langle k| \sum_{k_1,k_2} S_{k_1 k_2} |k_1\rangle\langle k_2| - \sum_{k_1,k_2} S_{k_1 k_2} |k_1\rangle\langle k_2| \sum_k I_S^{*(k)} |k\rangle\langle k|$$
$$= \sum_{k_1,k_2} S_{k_1 k_2} \left(I_S^{*(k_1)} - I_S^{*(k_2)}\right) |k_1\rangle\langle k_2| = 0. \quad (8)$$

Noting that $0 \equiv \sum_{k_1,k_2} 0 |k_1\rangle\langle k_2|$, from Eq. (8) we can see that

$$S_{k_1 k_2} \left(I_S^{*(k_1)} - I_S^{*(k_2)}\right) = \hat{0} \quad (9)$$

for arbitrary $k_1$ and $k_2$. Obviously, conditions (7) and (9) are equivalent. Thus, to satisfy LGKS Eq. (3), a COM has to commute with both $\hat{H}_S$ and $\hat{H}_{SR}$; conversely, any operator that commutes with $\hat{H}_S$ and $\hat{H}_{SR}$ is a COM.

It can be shown that Eqs. (3) and (4) ensure that the first and the second laws of thermodynamics are satisfied [6,23,26,28]. Usually, it is assumed that if there exists a COM of the system, the zeroth law of thermodynamics is violated. It implies that there are many stationary states of the system. Below we show how to construct these stationary states.

## III. SUBSPACES GENERATED BY SYSTEM-RESERVOIR INTERACTION AND CONSTANTS OF MOTION

If in the basis vectors $|k_s\rangle$ the matrix $\hat{S}$ defined by Eq. (5) has a block-diagonal form, then the whole space of the system states is a direct sum of subspaces corresponding to blocks of the matrix $\hat{S}$. If the initial system state belongs to one of such a subspace, then the system does not leave this subspace during the evolution. Indeed, using Eqs. (3) and (4) for diagonal and non-diagonal elements of the density matrix, we obtain

$$\dot{\rho}_S^{(k_1 k_1)} = \sum_{k_2=1}^N \gamma_{k_2 k_1} \left|S_{k_2 k_1}\right|^2 \rho_S^{(k_2 k_2)} - \rho_S^{(k_1 k_1)} \sum_{k_2=1}^N \gamma_{k_1 k_2} \left|S_{k_1 k_2}\right|^2 \quad (10)$$

$$\dot{\rho}_S^{(k_1 k_2)} = -\frac{1}{2} \sum_{k=1}^N \left(\gamma_{k k_1} \left|S_{k k_1}\right|^2 + \gamma_{k k_2} \left|S_{k k_2}\right|^2\right) \rho_S^{(k_1 k_2)} \quad (11)$$

where $\gamma_{k_1 k_2} = \lambda^2 G\left(\omega_{k_1} - \omega_{k_2}\right)$. From Eq. (11) one can see that any non-diagonal elements $\rho_S^{(k_1 k_2)}$ decays exponentially and does not interact with other elements. Equation (10) shows that diagonal elements $\rho_S^{(k_1 k_1)}$ interact only with other diagonal elements $\rho_S^{(k_2 k_2)}$ for which $S_{k_2 k_1} \neq 0$. This means that only intra-subsystem transitions that are determined by matrix elements related to a given subspace are possible. Thus, it is the form of the matrix of the operator $\hat{S}$ that determines the subspaces, in which the system evolves.

In 1937, Krylov [39] developed a special algorithm to construct the subspaces generated by an operator $\hat{S}$. A direct application of this algorithm, however, is not suitable for our purpose, because it includes the transition to new basis vectors. Since the LGKS equation implies the use of the basis vectors $|k_s\rangle$ of $\hat{H}_S$, then to reveal the block-diagonal form of the operator $\hat{S}$, we can only rearrange these vectors. Below, we modify the Krylov procedure in a way that the same basis vectors can be retained. This modification rearranges the basis vectors for the matrix of the operator $\hat{S}$ making it block-diagonal.

To construct the first subspace, we have to find the set of the basis vectors $B_1 = \{|k_i\rangle\}_1$, forming the first block of the matrix $\hat{S}$. The set $B_1$ should be constructed in a way that if some basis vector $|k_i\rangle$ belongs to $B_1$, then $S_{k_i k_j} = 0$ for any $|k_j\rangle \notin B_1$. The number of vectors in $B_1$ we denote as $N_1 \leq N$, where $N$ is the dimension of the whole space. We need to renumber the basis vector to place the vectors of $B_1$ at the beginning of the basis. This creates the first block in the upper-left corner of the matrix $S_{k_j k_i}$. Then, we have to repeat this procedure for the remaining basis vectors to create the next block and keep doing this until the whole matrix becomes block-diagonal.

To implement this recursive procedure, we start with some eigenvector $|k_1\rangle$ of the Hamiltonian $\hat{H}_S$ and construct the vector $\hat{S}|k_1\rangle$. Since $\hat{H}_S$ and $\hat{S}$ do not commute, the vector $\hat{S}|k_1\rangle$ may not be an eigenvector of $\hat{H}_S$. In this case, $\hat{S}|k_1\rangle$ can be represented as $\hat{S}|k_1\rangle = \sum_{i=1}^{n_1 < N} S_{k_i k_1} |k_i\rangle$ with $S_{k_i k_1} \neq 0$. This sum is a linear combination of $n_1$ basis vectors corresponding to non-zero elements in the $k_1$-th column of the matrix $S_{k_i k_1}$. These $n_1 \leq N$ vectors form the set $B_1$. On the next step, we decompose each vector $|k_i\rangle$ of the set $B_1$ as $\hat{S}|k_i\rangle = \sum_{j=1}^{N} S_{k_j k_i} |k_j\rangle$. If in the decompositions, vectors $|k_j\rangle$ that do not belong to $B_1$ arise, then we should add them to $B_1$. The procedure should be repeated until on some step no new vectors arise in the decompositions. This completes the construction of the set $B_1$ containing $N_1$ vectors. Then, we should rearrange the basis vectors in a way that all vectors of $B_1$ take the first $N_1$ positions in the basis. As a result, in the upper-left-corner of the matrix $S_{k_j k_i}$, we form a diagonal block.

If $N_1 = N$, then the dimension of this block is equal to the dimension of the whole space. If $N_1 < N$, then the above procedure should be repeated with the vector $|k_{N_1+1}\rangle$ in the rearranged basis. We obtain the next block and so on.

This construction ensures that in the rearranged basis, the matrix of the operator $\hat{S}$ has a block-diagonal form. By construction, this decomposition of the system state space is invariant.

Now we show that the constructed subspaces determine all possible COMs. First, we show that for any subspace $C_{l_0}$, the operator

$$\hat{I}_S = I_S^{(l_0)} \sum_{k_i^{(l_0)} \in B_{l_0}} |k_i^{(l_0)}\rangle\langle k_i^{(l_0)}| = I_S^{(l_0)} \hat{P}_{l_0},$$

where $I_S^{(l_0)}$ is some $c$-number, is a COM. For this, we have to prove that $\hat{I}_S$ commutes with both $\hat{H}_S$ and $\hat{S}$. Note that $\hat{P}_{l_0}$ is the projection operator onto the $l_0$-th subspace, i.e., it is a unitary operator in $B_{l_0}$ and is zero in other subspaces. Because $\hat{I}_S$ is diagonal in the basis of eigenvectors of $\hat{H}_S$, then $\left[\hat{I}_S, \hat{H}_S\right] = 0$.

Next, in the rearranged basis, the operator $\hat{S}$ is block-diagonal, therefore, $\hat{S} = \sum_l \sum_{k_{i_1}^{(l)}, k_{i_2}^{(l)} \in B_l} S_{i_1, i_2} |k_{i_1}^{(l)}\rangle\langle k_{i_2}^{(l)}|$. Then

$$\left[\hat{I}_S, \hat{S}\right] = I_S^{(l_0)} \left[\hat{1}_{l_0}, \sum_{k_{i_1}^{(l_0)}, k_{i_2}^{(l_0)} \in B_{l_0}} S_{i_1, i_2} |k_{i_1}^{(l_0)}\rangle\langle k_{i_2}^{(l_0)}|\right] = 0. \tag{12}$$

Thus, $\hat{I}_S = I_S^{(l_0)} \hat{P}_{l_0}$ is a COM. As a consequence, any operator, which can be decomposed as

$$\hat{I}_S = \sum_l I_S^{(l)} \hat{P}_l, \tag{13}$$

where $I_S^{(l)}$ are arbitrary $c$-numbers, which are fixed for a given subspace, is also a COM as a linear combination of COMs.

Now, we show that there are no other COMs apart from those having from (13). Let us assume the contrary: a COM, $\hat{I}_S^*$, which cannot be expressed in form (13), exists. Since $\hat{I}_S^*$ is a COM, it commutes with $\hat{H}_S$ and $\hat{S}$. Because the operator $\hat{I}_S^*$ commutes with $\hat{H}_S$, it is diagonal in the basis $|k\rangle$ of $\hat{H}_S$ eigenvectors. Next, due to commutativity of $\hat{I}_S^*$ and $\hat{S}$, Eq. (9) must be satisfied. According to our assumption, $\hat{I}_S^*$ cannot be presented in form (13). Hence, *in some subspace*, vectors having different eigenvalues, say $I_S^{*(k_j)} \neq I_S^{*(k_i)}$, must exist. From Eq. (9) it follows that if vectors $|k_i\rangle$ and $|k_j\rangle$ have different eigenvalues, then $S_{k_i k_j} = 0$. This means that it is possible to combine the vectors with identical eigenvalues into new subspaces so that the operator $\hat{S}$ takes a block-diagonal form inside the initial block. This contradicts the fact that invariant subspaces constructed above cannot be divided into invariant subspaces of lower dimensions. Therefore, there are no COMs apart from those that have form (13). Moreover, this means that during the system evolution, no values of COMs change. Thus, this division corresponds to the basis family of COMs. By construction, starting at any point in an invariant subspace, the system should visit all points of this subspace.

## IV. STATIONARY SOLUTIONS OF THE MASTER EQUATION AND CONSTANTS OF MOTION

Now, we can find the stationary solution of the master equation. For Eqs. (3) and (4) along with Kubo-Martin-Schwinger condition, the stationary solution is the Gibbs distribution:

$$\hat{\rho}_S^{th} = \exp(-\hat{H}_S / kT) / \text{Tr} \exp(-\hat{H}_S / kT). \tag{14}$$

This can be verified by the direct substitution of Eq. (14) into Eq. (3). However, this stationary solution may not be unique. If there are invariant subspaces, then the Gibbs distribution over the states of a given invariant subspace is also a stationary solution. Then, any state of the form

$$\hat{\rho}_S^{st} = \sum_j \lambda_j \frac{\exp(-\hat{P}_j \hat{H} \hat{P}_j / kT)}{\text{Tr} \exp(-\hat{P}_j \hat{H} \hat{P}_j / kT)}, \quad \sum_j \lambda_j = 1, \quad 0 \leq \lambda_j \leq 1, \tag{15}$$

is stationary. Because LGKS Eq. (3) conserves the trace and $\left\{ \left| k_i^{(j)} \right\rangle \right\}_{i=\sum_{l=1}^{j-1} N_l + 1, \sum_{l=1}^{j-1} N_l + N_j}$ are invariant subspaces, the value $\text{Tr}_j \rho_S(t)$ does not change in time. Therefore, $\lambda_{N_j} = \text{Tr}_{N_j} \hat{\rho}_S^{st} = \text{Tr}_{N_j} \hat{\rho}_S(0)$. Thus, in

each invariant subspace, the system state evolves to the Gibbs distribution over the states of the subspace with the partition function $\text{Tr}\exp(-\hat{P}_{N_j}\hat{H}\hat{P}_{N_j}/kT)$. In each invariant subspaces, there are no non-trivial COMs. As shown in Refs. [28-30], this condition is necessary and sufficient for the uniqueness of a stationary solution. Thus, Eq. (15) determines all possible stationary solutions.

In a particular case, when the operator $\hat{S}$ commutes with the Hamiltonian $\hat{H}_S$, *all* the nondiagonal elements of $\hat{S}$ in the basis of the eigenvectors of $\hat{H}_S$ are equal to zero, and each subset $C_l$ includes only one eigenstate ($N_j = 1$ for each $j$). Then, any operator that is diagonal in the basis of the eigenstates of the Hamiltonian $\hat{H}_S$ is a COM. In particular, the Hamiltonian $\hat{H}_S$ itself is a COM; therefore, the energy of the system does not chane in time. The system does not have the Gibbs distribution, and the distribution depends on the initial state. An example of such a situation is a dephasing reservoir (see Ref. [27]).

## V. EXAMPLE: INTERACTING TWO-LEVEL SYSTEMS

To illustrate the results obtained above we apply the depeloped procedure to a system of two interacting two-level subsystems (TLSs) which relax into a dephasing reservoir. We begin with considering non-interacting TLSs.

### A. Non-interacting TLSs

Suppose that the transition frequencies of TLSs are $\omega_1$ and $\omega_2$, we denote excited and ground states as $|e_i\rangle$ and $|g_i\rangle$ and the transition oparators between excited and ground states of each TSL as $\hat{\sigma}_i$, $i = 1, 2$. The total Hamiltonian of the system is

$$\hat{H}_S = \hat{H}_1 + \hat{H}_2 = \hbar\omega_1\hat{\sigma}_1^\dagger\hat{\sigma}_1 + \hbar\omega_2\hat{\sigma}_2^\dagger\hat{\sigma}_2, \tag{16}$$

with eigenstates $|e_1, e_2\rangle, |e_1, g_2\rangle, |g_1, e_2\rangle, |g_1, g_2\rangle$ and eigenvalues $\omega_1 + \omega_2, \omega_1, \omega_2, 0$.

Suppose that TLSs interact with the reservoir desribed by the Hamiltonian:

$$\hat{H}_R = \hbar\sum_k \omega_k \hat{a}_k^\dagger \hat{a}_k, \tag{17}$$

where $\omega_k$ is the frequency of the *k*-th reservoir mode, and the interaction Hamiltonian is

$$\hat{H}_{SR} = \hbar\sum_k \gamma_1^k \hat{\sigma}_1^z \left(\hat{a}_k^\dagger + \hat{a}_k\right) + \hbar\sum_k \gamma_2^k \hat{\sigma}_2^z \left(\hat{a}_k^\dagger + \hat{a}_k\right). \tag{18}$$

where $\gamma_1^k$ and $\gamma_2^k$ are the interaction constants between the first and the second TLSs and the $k$-th reservoir mode, respectively, and $\hat{\sigma}_i^z = [\hat{\sigma}_i^\dagger, \hat{\sigma}_i]$ is the operator of the population inversion of the $i$-th TLS. For simplicity, we assume that $\gamma_2^k = a\gamma_1^k$, where the constant $a$ does not depend on $k$. Then,

$$\hat{H}_{SR} = \hbar \sum_k \gamma_1^k \left(\hat{\sigma}_1^z + a\hat{\sigma}_2^z\right)\left(\hat{a}_k^\dagger + \hat{a}_k\right) = \hbar \lambda \hat{S}\hat{R}, \tag{19}$$

where $\lambda = \max\{\gamma_1^k\}$, $\hat{R} = \sum_k \dfrac{\gamma_1^k}{\max(\gamma_1^k)}\left(\hat{a}_k^\dagger + \hat{a}_k\right)$, and $\hat{S} = \hat{\sigma}_1^z + a\hat{\sigma}_2^z$. Such reservoir describes phase relaxation of the system. Indeed, the operator $\hat{S} = \hat{\sigma}_1^z + a\hat{\sigma}_2^z$ commutes with the system Hamiltonian $\hat{H}_S$ and the energy of the system is conserved; thus, the reservoir is purely dephasing. According to Sec. IV, in this case, each invariant subspace consists of only one system eigenstate.

To show this explicitily, we follow the procedure developed in Sec. III. Acting by the operator $\hat{S}$ on the eigenstates of $\hat{H}_S$, we obtain

$$\hat{S}|e_1, e_2\rangle = (1+a)|e_1, e_2\rangle, \; \hat{S}|g_1, g_2\rangle = -(1+a)|g_1, g_2\rangle, \tag{20}$$

$$\hat{S}|g_1, e_2\rangle = (-1+a)|g_1, e_2\rangle, \; \hat{S}|e_1, g_2\rangle = (1-a)|e_1, g_2\rangle. \tag{21}$$

In the action on each eigenvector, no new eigenvectors appear. Thus, each eigenvector forms invariant subspace with the dimension one.

The corresponding COMs are projections over each invarinat subspaces, namely, $\hat{P}_1 = |e_1, e_2\rangle\langle e_1, e_2|$, $\hat{P}_2 = |g_1, g_2\rangle\langle g_1, g_2|$, $\hat{P}_3 = |e_1, g_2\rangle\langle e_1, g_2|$, and $\hat{P}_4 = |g_1, e_2\rangle\langle g_1, e_2|$. These COMs are basis COMs, and any linear combination of them is also a COM. Since $\sum_i \hat{P}_i = \hat{1}$, out of four COMs, only three are linear independent.

In this simple example, we can construct linear combinations that have clear physical meanings. The first one is

$$\begin{aligned}2\hat{P}_1 + 1\hat{P}_3 + 1\hat{P}_4 + 0\hat{P}_2 &= 2|e_1, e_2\rangle\langle e_1, e_2| + 1|e_1, g_2\rangle\langle e_1, g_2| + \\&+ 1|g_1, e_2\rangle\langle g_1, e_2| + 0|g_1, g_2\rangle\langle g_1, g_2| = \\&= \left(|e_1, e_2\rangle\langle e_1, e_2| + |e_1, g_2\rangle\langle e_1, g_2|\right) + \left(|e_1, e_2\rangle\langle e_1, e_2| + |g_1, e_2\rangle\langle g_1, e_2|\right) = \\&= \hat{\sigma}_1^\dagger\hat{\sigma}_1 + \hat{\sigma}_2^\dagger\hat{\sigma}_2\end{aligned} \tag{22}$$

This operator discribes the number of excitations in the system. Indeed, $\hat{P}_1$ corresponds to the state in which both TLSs are in the excited states and there are two excitations in the system, $\hat{P}_2$ corresponds to the state in which both TLSs are in the ground state and there are no excitations in the system. $\hat{P}_3$ and $\hat{P}_4$ correspond to the subspaces in which only one of TLSs is excited and there is only one excitation. Thus, the operator $2\hat{P}_1 + 1\hat{P}_3 + 1\hat{P}_4 + 0\hat{P}_2 = \hat{\sigma}_1^\dagger \hat{\sigma}_1 + \hat{\sigma}_2^\dagger \hat{\sigma}_2$ has the eigenvalue which is the number of excitations.

The second linear combination is

$$2\hat{P}_1 - 2\hat{P}_2 + 0\hat{P}_3 + 0\hat{P}_4 = 2|e_1,e_2\rangle\langle e_1,e_2| - 2|g_1,g_2\rangle\langle g_1,g_2| = \\
= (|e_1,e_2\rangle\langle e_1,e_2| + |e_1,g_2\rangle\langle e_1,g_2|) - (|g_1,e_2\rangle\langle g_1,e_2| + |g_1,g_2\rangle\langle g_1,g_2|) + \\
+ (|e_1,e_2\rangle\langle e_1,e_2| + |g_1,e_2\rangle\langle g_1,e_2|) - (|e_1,g_2\rangle\langle e_1,g_2| + |g_1,g_2\rangle\langle g_1,g_2|) = \\
= \hat{\sigma}_1^z + \hat{\sigma}_2^z. \quad (23)$$

The operator $\hat{\sigma}_1^z + \hat{\sigma}_2^z$ discribes the total population inversion of the system. Indeed, in the first subspace, the state $|e_1,e_2\rangle$ corresponds to two excited TLSs with the population inversion of 2, in the second subspace, the state is $|g_1,g_2\rangle$ and the population inversion is –2, in subsapces $|e_1,g_2\rangle$ and $|g_1,e_2\rangle$, the population inversion is zero.

The third linear combination of basis COMs is the total energy of the system:

$$(\omega_1+\omega_2)\hat{P}_1 + \omega_1\hat{P}_2 + \omega_2\hat{P}_3 + 0\hat{P}_4 = (\omega_1+\omega_2)|e_1,e_2\rangle\langle e_1,e_2| + \\
+ \omega_1|e_1,g_2\rangle\langle e_1,g_2| + \omega_2|g_1,e_2\rangle\langle g_1,e_2| + 0|g_1,g_2\rangle\langle g_1,g_2| = \\
= \omega_1(|e_1,e_2\rangle\langle e_1,e_2| + |e_1,g_2\rangle\langle e_1,g_2|) + \omega_2(|e_1,e_2\rangle\langle e_1,e_2| + |g_1,e_2\rangle\langle g_1,e_2|) = \\
= \omega_1\hat{\sigma}_1^\dagger\hat{\sigma}_1 + \omega_2\hat{\sigma}_2^\dagger\hat{\sigma}_2 = \hat{H}_S \quad (24)$$

Note that the total energy of the system as well as energies of each TLSs are conserved. For this reason, the reservoir with Hamiltonian (17) and interaction (18) may be called dephasinfg.

These three COMs, the number of system excitation, the total population inversion, and the total system energy fully characterize the final state of the system.

### B. Interacting TLSs

Now suppose that there is dipole-dipole interaction between TLSs so that the interaction between them is descibed by the Hamiltionian $\hat{V} = (\hat{\mathbf{d}}_1\hat{\mathbf{d}}_2 - 3(\hat{\mathbf{d}}_1\mathbf{n})(\hat{\mathbf{d}}_2\mathbf{n}))/r^3$, where $r$ is the distance between TLSs, $\mathbf{n}$ is the normal vector directed from one TLS to another. Using the expression for TLS dipole moment, $\hat{\mathbf{d}}_i = \mathbf{d}_i^{eg}(\hat{\sigma}_i + \hat{\sigma}_i^\dagger)$ ($\mathbf{d}_i^{eg}$ is the matrix element of the dipole transtion), the interaction Hamiltonian in the rotating-wave approximation can be rewritten as $\hat{V} = \hbar\Omega_R(\sigma_1^\dagger\hat{\sigma}_2 + \hat{\sigma}_2^\dagger\hat{\sigma}_1)$, where $\Omega_R = \left(\mathbf{d}_1^{eg}\mathbf{d}_2^{eg} - 3\left(\mathbf{d}_1^{eg}\mathbf{n}\right)\left(\mathbf{d}_2^{eg}\mathbf{n}\right)\right)/r^3$ is the Rabi constant of the interaction. The Hamiltonian of the system may be written as

$$\hat{H}_S = \hat{H}_1 + \hat{H}_2 + \hat{V} = \hbar\omega_1 \hat{\sigma}_1^\dagger \hat{\sigma}_1 + \hbar\omega_2 \hat{\sigma}_2^\dagger \hat{\sigma}_2 + \hbar\Omega_R\left(\hat{\sigma}_1^\dagger \hat{\sigma}_2 + \hat{\sigma}_2^\dagger \hat{\sigma}_1\right). \tag{25}$$

Eigenstates of $\hat{H}_S$ are

$$\begin{aligned}|\psi_1\rangle = |e_1, e_2\rangle, \quad |\psi_2\rangle = |g_1, g_2\rangle, \\ |\psi_3\rangle = \cos\varphi|e_1, g_2\rangle + \sin\varphi|g_1, e_2\rangle, \quad |\psi_4\rangle = -\sin\varphi|e_1, g_2\rangle + \cos\varphi|g_1, e_2\rangle,\end{aligned} \tag{26}$$

where

$$\varphi = \tan^{-1}\left(\left(\sqrt{\Delta\omega^2/4 + \Omega_R^2} - \Delta\omega/2\right)/\Omega_R\right). \tag{27}$$

The eigenvalues of eigenstates (26) are

$$E_1 = \omega_1 + \omega_2, \quad E_2 = 0, \quad E_{3,4} = (\omega_1 + \omega_2)/2 \pm \sqrt{\Delta\omega^2/4 + \Omega_R^2}. \tag{28}$$

Note that the interaction between TLSs results in mixing of states $|e_1, g_2\rangle$ and $|g_1, e_2\rangle$ [see Eq. (26)].

Now we follow the procedure developed in Sec. III. Equation (20) holds as before, because the first two eigenvectors, $|\psi_1\rangle$ and $|\psi_2\rangle$, are equal to $|e_1, e_2\rangle$ and $|g_1, g_2\rangle$, respectively. Since the interaction operator $\hat{V}$ mixes the states $|e_1, g_2\rangle$ and $|g_1, e_2\rangle$, instead of Eq. (21) the action of the operator $\hat{S}$ on the states $|\psi_3\rangle$ and $|\psi_4\rangle$ should be considered. As a result, we have

$$\hat{S}|\psi_3\rangle = \left(|\alpha_3|^2 - |\beta_3|^2\right)(1-a)|\psi_3\rangle + \left(\alpha_3\alpha_4^* - \beta_3\beta_4^*\right)(1-a)|\psi_4\rangle. \tag{29}$$

We can see that $|\psi_3\rangle$ is no longer an eigenvector of $\hat{S}$. The result of the action of $\hat{S}$ on $|\psi_3\rangle$, in addition to $|\psi_3\rangle$, contains another basis vector, $|\psi_4\rangle$. Now, we should act by the operator $\hat{S}$ on this vector:

$$\hat{S}|\psi_4\rangle = \left(|\alpha_4|^2 - |\beta_4|^2\right)(1-a)|\psi_4\rangle + \left(\alpha_4\alpha_3^* - \beta_4\beta_3^*\right)(1-a)|\psi_3\rangle. \tag{30}$$

There are no new basis vectors in Eq. (30). Thus, the subspace spanned by the basis vectors $|\psi_3\rangle$ and $|\psi_4\rangle$ is an invariant subspace with the dimension of two. Thus, number of invariant subsapces is reduced from four to three. The projection operator on the invariant subspace spanned by the basis vectors $|\psi_3\rangle$ and $|\psi_4\rangle$ is

$$\begin{aligned}\hat{\tilde{P}} &= |\psi_3\rangle\langle\psi_3| + |\psi_4\rangle\langle\psi_4| = \left(\cos\varphi|e_1,g_2\rangle + \sin\varphi|g_1,e_2\rangle\right)\left(\cos\varphi\langle e_1,g_2| + \sin\varphi\langle g_1,e_2|\right) + \\ &+ \left(-\sin\varphi|e_1,g_2\rangle + \cos\varphi|g_1,e_2\rangle\right)\left(-\sin\varphi\langle e_1,g_2| + \cos\varphi\langle g_1,e_2|\right) = \\ &= \left(\cos^2\varphi + \sin^2\varphi\right)|e_1,g_2\rangle\langle e_1,g_2| + \left(\sin^2\varphi + \cos^2\varphi\right)|g_1,e_2\rangle\langle g_1,e_2| + \\ &+ \left(\cos\varphi\sin\varphi - \sin\varphi\cos\varphi\right)|e_1,g_2\rangle\langle g_1,e_2| + \left(\sin\varphi\cos\varphi - \cos\varphi\sin\varphi\right)|g_1,e_2\rangle\langle e_1,g_2| = \\ &= |e_1,g_2\rangle\langle e_1,g_2| + |g_1,e_2\rangle\langle g_1,e_2| = \hat{P}_3 + \hat{P}_4\end{aligned} \quad (31)$$

It should be emphasized that in this case neither $\hat{P}_3$ nor $\hat{P}_4$ is a COM, but their combination $\hat{\tilde{P}}$ is.

In this case, there are two linear independent COMs. The linear combinations that have physical meaning are the number of excitation, $\hat{\sigma}_1^\dagger\hat{\sigma}_1 + \hat{\sigma}_2^\dagger\hat{\sigma}_2 = 2\hat{P}_1 + 1\hat{\tilde{P}} + 0\hat{P}_2$, and the total population inversion, $2\hat{P}_1 - 2\hat{P}_2 = \hat{\sigma}_1^z + \hat{\sigma}_2^z$. Due to the interaction between TLSs, the system Hamiltonian is no longer a COM. This means that the reservoir ceases to be purely dephasing and results in energy relaxation in the invariant subspace spanned by the basis vectors $|\psi_3\rangle$ and $|\psi_4\rangle$. It remains dephasing, however, in the subspaces with vectors $|\psi_1\rangle$ and $|\psi_2\rangle$.

Using the obtained COMs and Eq. (15), we may write possible stationary solutions of the corresponding LGKS equation:

$$\hat{\rho}_S^{st} = \lambda_1|\psi_1\rangle\langle\psi_1| + \lambda_2|\psi_2\rangle\langle\psi_2| + \frac{\lambda}{1+\exp\left(-\frac{E_3-E_4}{kT}\right)}\left(|\psi_4\rangle\langle\psi_4| + \exp\left(-\frac{E_3-E_4}{kT}\right)|\psi_3\rangle\langle\psi_3|\right) \quad (32)$$

where $\lambda$, $\lambda_1$, and $\lambda_2$ are detremined by the initial density matrix $\hat{\rho}(0)$:

$$\lambda_1 = \rho_{11}(0),\ \lambda_1 = \rho_{11}(0),\ \lambda = \rho_{33}(0) + \rho_{44}(0) \quad (33)$$

Note that in the invariant subspaces with dimension 1, the stationary and initial states are the same. In the invariant subsapce with dimension 2, the stationary solution is the Gibbs distribution.

In Fig. 1, the dependences of the matrix elements $\rho_{33}(t)$ and $\rho_{44}(t)$ on time obtained by computer simulation of the Eq. (10) are shown. One can see that they indeed converge to the Gibbs distribution.

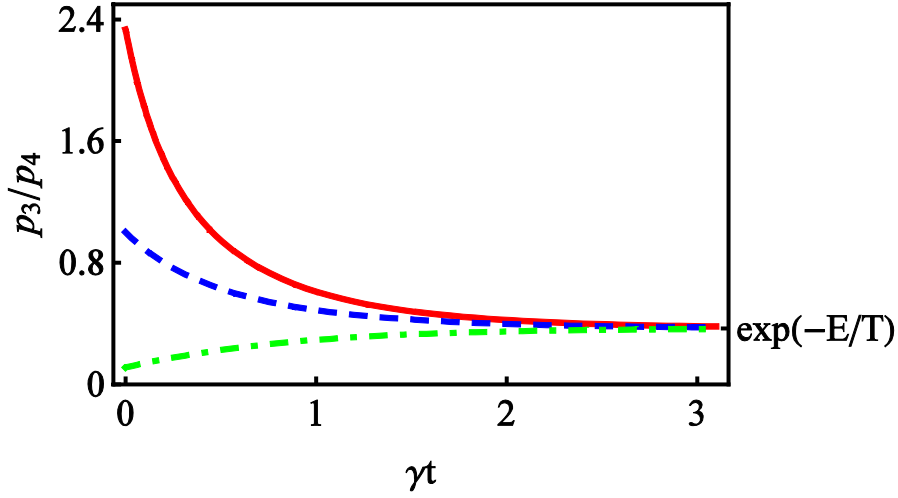

Fig. 1. The dependence of the diagonal matrix elements of the density matrix $p_3 = \rho_{33}$ and $p_4 = \rho_{44}$ on time obtained from LGKS Eq. (10) for different initial condition: $p_3(0) = 0.7$ and $p_4(0) = 0.3$ (the red solid line), $p_3(0) = 0.7$ and $p_4(0) = 0.3$ (the blue dashed line), $p_3(0) = 0.1$ and $p_4(0) = 0.9$ (the green dot-dashed line); $E = E_3 - E_4 = T = 1$, $\gamma_{34} = 1$, $\gamma_{43} = \gamma_{34} \exp(-E/T)$, $t$ is expressed in the units of $\gamma_{34}$.

## VI. CONCLUSION

In this work, we consider stationary states of an open quantum system interacting with a thermal reservoir in a system that has COMs. We show that stationary states retain the memory of the initial state of the system. To be specific, using the basis of eigenfunctions of the system Hamiltonian $H_S$, we have shown that the Hamiltonian of the interaction between the system and the reservoir $H_{SR}$ determines the splitting of the space of system states into a set of subspaces. In each of the subspaces, the system behaves as if there are no COMs. This means that, if the initial state of the system belongs to one of these subspaces, the system evolves inside this subspace reaching the Gibbs distribution after thermalization. Hence, each such an invariant subspace can be linked to a COM by assigning some eigenvalue to this COM (say, unity, in one invariant subspace and zeros in the others). Consequently, each subspace determines its own COM that has a fixed eigenvalue in this subspace and zeros in others. If there are $N$ subspaces, then it is possible to define $N$–1 COMs because in each subspace, COMs with identical values are trivial and do not lead to non-uniqueness of the stationary state. Thus, the algorithm developed in the paper allows one to find *all* invariant subspaces and *all* COMs.

The eigenvalues of existing COMs determine neither the stationary state in each subspace nor the stationary state of the whole system. In any subspace, the Gibbs distribution is determined by the temperature of the reservoir and by the set of eigenfunctions of $H_S$, which construct this subspace. To find the stationary state of the whole system, one must know the initial state of the system. The projection of this state onto subspaces provides the weight factors for Gibbs distributions characterizing each subspace. The weight factors determine the corresponding stationary state of the whole system as a weighted sum of the Gibbs distributions over the subspaces.

Thus, as an open quantum system with COMs interacting with a reservoir evolves, it reaches one of many possible stationary states. Though this state is thermalized with the temperature of the reservoir, it is determined by the initial state of the system.

A.A.L acknowledges the support of the National Science Foundation under Grants No. DMR-1312707.

[1] P. A. Camati, J. P. Peterson, T. B. Batalhao, K. Micadei, A. M. Souza, R. S. Sarthour, I. S. Oliveira, and R. M. Serra, Phys. Rev. Lett. **117**, 240502 (2016).
[2] J. Gemmer, M. Michel, and G. Mahler, *Quantum thermodynamcis - emergence of thermodynamic behavior within composite quantum systems*, 2nd. ed. (Springer, Berlin, 2010).
[3] J. Goold, M. Huber, A. Riera, L. del Rio, and P. Skrzypczyk, J. Phys. A: Math Theor. **49**, 143001 (2016).
[4] C. Jarzynski, Phys. Rev. X **7**, 011008 (2017).
[5] J. Koski, T. Sagawa, O. Saira, Y. Yoon, A. Kutvonen, P. Solinas, M. Möttönen, T. Ala-Nissila, and J. Pekola, Nat. Phys. **9**, 644 (2013).
[6] R. Kosloff, Entropy **15**, 2100 (2013).
[7] M. Lostaglio, D. Jennings, and T. Rudolph, Nat. Commun. **6**, 2015).
[8] E. A. Martinez and J. P. Paz, Phys. Rev. Lett. **110**, 130406 (2013).
[9] J. M. Parrondo, J. M. Horowitz, and T. Sagawa, Nat. Phys. **11**, 131 (2015).
[10] U. Seifert, Rep. Prog. Phys. **75**, 126001 (2012).
[11] J. Shiner, *Entropy and Entropy Generation: Fundamentals and Applications* (Springer Science & Business Media, 2001).
[12] P. Skrzypczyk, A. J. Short, and S. Popescu, Nat. Commun. **5**, 4185 (2014).
[13] S. Toyabe, T. Sagawa, M. Ueda, E. Muneyuki, and M. Sano, Nat. Phys. **6**, 988 (2010).
[14] O. Gühne and G. Tóth, Physics Reports **474**, 1 (2009).
[15] J. Preskill, *Lecture notes for physics 229: Quantum information and computation* (California Institute of Technology, 1998).
[16] T. D. Ladd, F. Jelezko, R. Laflamme, Y. Nakamura, C. Monroe, and J. L. O'Brien, Nature **464**, 45 (2010).

[17] M. Eisaman, J. Fan, A. Migdall, and S. V. Polyakov, Review of scientific instruments **82**, 071101 (2011).
[18] C. J. Chunnilall, I. P. Degiovanni, S. Kück, I. Müller, and A. G. Sinclair, Opt. Eng. **53**, 081910 (2014).
[19] D. J. Bergman and M. I. Stockman, Phys. Rev. Lett. **90**, 027402 (2003).
[20] M. Noginov, G. Zhu, A. Belgrave, R. Bakker, V. Shalaev, E. Narimanov, S. Stout, E. Herz, T. Suteewong, and U. Wiesner, Nature **460**, 1110 (2009).
[21] A. H. Schokker and A. F. Koenderink, Optica **3**, 686 (2016).
[22] W. Zhou, M. Dridi, J. Y. Suh, C. H. Kim, M. R. Wasielewski, G. C. Schatz, and T. W. Odom, Nat. Nano **8**, 506 (2013).
[23] E. B. Davies, Commun. Math. Phys. **39**, 91 (1974).
[24] V. Gorini, A. Kossakowski, and E. C. G. Sudarshan, Journal of Mathematical Physics **17**, 821 (1976).
[25] G. Lindblad, Communications in mathematical Physics **48**, 119 (1976).
[26] H.-P. Breuer and F. Petruccione, *The theory of open quantum systems* (Oxford University Press on Demand, 2002).
[27] A. Rivas and S. F. Huelga, *Open Quantum Systems* (Springer, 2012).
[28] H. Spohn and J. L. Lebowitz, Adv. Chem. Phys **38**, 109 (1978).
[29] A. Frigerio, Lett. Math. Phys. **2**, 79 (1977).
[30] H. Spohn, Rev. Mod. Phys. **52**, 569 (1980).
[31] H. J. Carmichael, *Statistical Methods in Quantum Optics 2: Non-Classical Fields* (Springer Science & Business Media, 2009).
[32] U. Weiss, *Quantum dissipative systems* (World scientific, 2012).
[33] B. Baumgartner and H. Narnhofer, J. Phys. A: Math Theor. **41**, 395303 (2008).
[34] B. Baumgartner, H. Narnhofer, and W. Thirring, J. Phys. A: Math Theor. **41**, 065201 (2008).
[35] F. Ticozzi, R. Lucchese, P. Cappellaro, and L. Viola, IEEE Trans. Autom. Control **57**, 1931 (2012).
[36] F. Ticozzi, S. G. Schirmer, and X. Wang, IEEE Trans. Autom. Control **55**, 2901 (2010).
[37] F. Ticozzi and L. Viola, IEEE Trans. Autom. Control **53**, 2048 (2008).
[38] S. Schirmer and X. Wang, Phys. Rev. A **81**, 062306 (2010).
[39] F. R. Gantmacher, *The theory of matrices* (AMS Chelsea, 1998).